# The Very Low 1-9 GeV/c Tertiary Beam Extension of the H8 Beam Line of CERN SPS

**I. Efthymiopoulos, A. Fabich**
*AB Department, CERN*

**Abstract**

The H8 beam line on the North Area of CERN SPS is a multi purpose experimental secondary particle beam able to transport particles of various types (hadrons, electrons, muons and ions) in a large momentum range from $10\,\mathrm{GeV/c}$ up to the top SPS energy (400 or $450\,\mathrm{GeV/c}$). During the 2003-2004 shutdown the beam line was modified in order to accommodate a new tertiary beam line, able to produce and transport particles in the momentum range of 1 to $9\,\mathrm{GeV/c}$. Such low momentum hadron and electron particle beams are requested by the LHC experiments in order to calibrate various sub-detector systems. The design and first performance results from this new tertiary beam are described here.

# 1 Introduction

The H8 beam line in the North Area of CERN was designed to deliver secondary and tertiary beams of various particle types (hadron, electron, muon, ions), as well as an attenuated primary proton beam, to fixed target experiments located at the EHN1 experimental hall. The momentum range of the beam line is quite large, from $10\,\mathrm{GeV/c}$ up to the maximum SPS momentum of $450\,\mathrm{GeV/c}$. Secondary beams are produced directly from the impinging protons from SPS at the T4 target, while using a secondary target located at about 130 m downstream of the T4 target tertiary beams can be produced. For both cases, a large spectrometer constructed of six MBN dipole magnets is used for the momentum definition. The maximum momentum acceptance of the beam line is $\pm 1.4\%$ in $\Delta p/p$.

Currently the ATLAS collaboration is the main user of H8 the beam line, performing tests of various detector prototype and production modules. The last ingredient for their calibrations is to verify the response of their detectors in the very low energy range i.e. down to 1 GeV/c, since experience from previous detectors indicate that understanding the response at low energies is crucial for the final detector performance and jet reconstruction [1].

Although the obvious place to perform such tests would be the East Hall at PS, due to the large size of the ATLAS detectors and the heavy installations involved (for example the cryogenics for the LArgon calorimeters), such an option was excluded. Thus the possibility of making such **V**ery **L**ow **E**nergy (**VLE**) beams of both hadrons (pions) and electrons in H8 was proposed. Using the existing tertiary mode is not possible for two reasons:

- for such low energies the required current in the MBN spectrometer magnets has to be very small ($\sim$ 3A), reaching the limit in terms of stability and precision of the power supplies.
- low energy pions transported at large distances (there are about $500\,\mathrm{m}$ between the secondary target and the ATLAS installation in H8A area) will decay before reaching the experiment.

To overcome these difficulties the secondary target has to be located nearby the experimental setup accompanied with a momentum selection station and focusing elements. Finally, another important constraint by the experiment was to have both the VLE and the existing secondary beams arriving at the same place at the experiment in order to allow a continuous calibration with the same setup.

The conceptual design of the VLE tertiary beam for hadron (pion) and electron beams in the range of $1-10\,\mathrm{GeV/c}$ of both signs, is shown in Figure 1.

Using a medium energy, high intensity secondary beam from T4 directed towards the secondary target (T48) located at about $40\,\mathrm{m}$ upstream of the experimental setup, particle rates of at least few hundred particles per spill at all energies would be possible. Two large acceptance quadrupoles are located nearby the target followed by a momentum selection and recombination station with four large acceptance bending magnets. The momentum acceptance is maximum $\pm 4.5\%$ which is still acceptable, given that for example the em-calorimeter resolution is $8\%$ at $1\,\mathrm{GeV/c}$. A spectrometer is also available around the fourth bending magnet to provide particle-by-particle momentum measurement with a resolution better than $1\%$. Finally, a set of focusing quadrupoles provides a beam spot at the experiment of about $5 \times 5\,\mathrm{cm}^2$. The design is completed with a dump next to the second bending magnet to stop the remnant of the secondary beam from reaching the experiment.

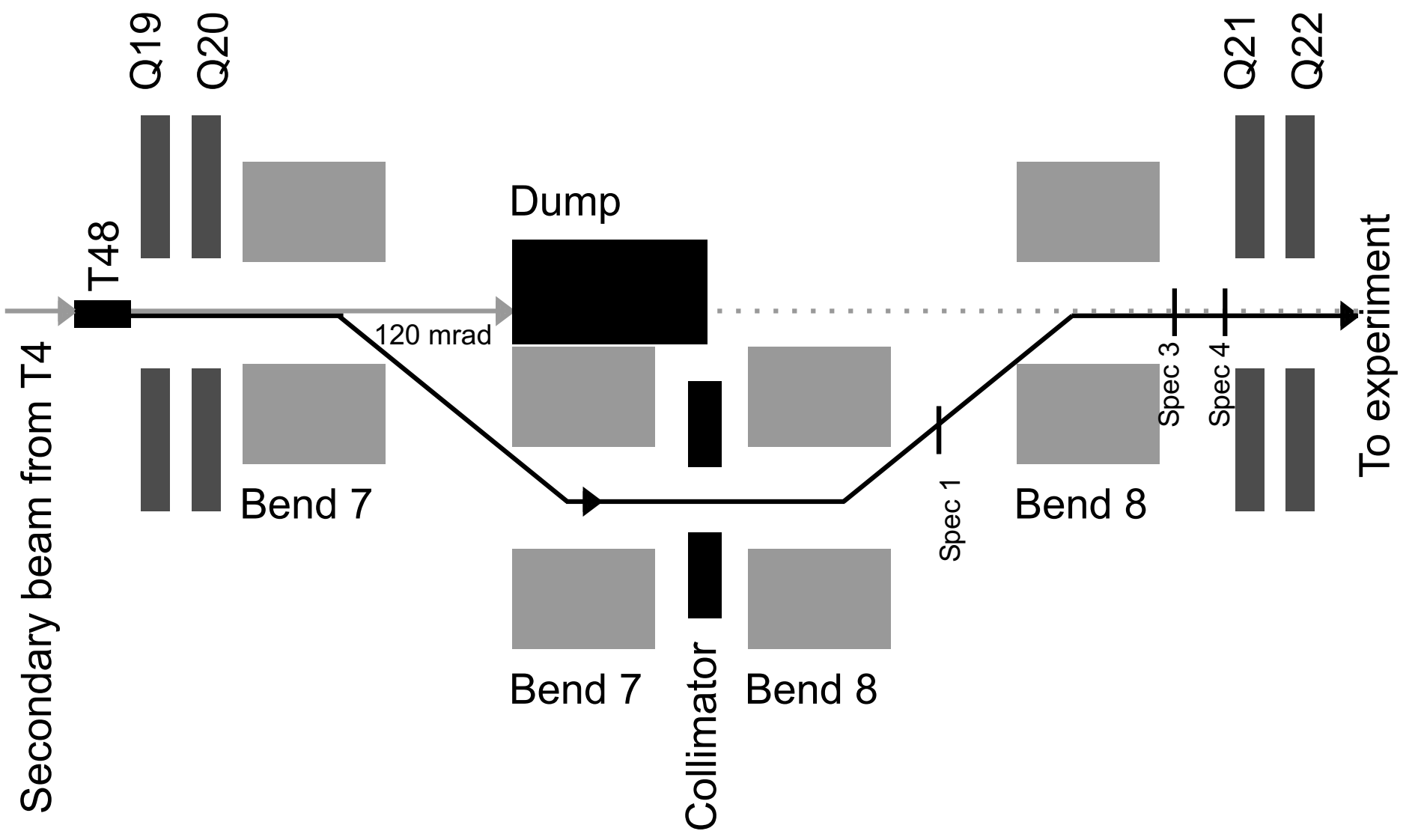

Figure 1: *Schematic layout of the VLE setup in H8 beam line.*

The design is such that the instrumentation and performance of all existing beams remains intact. The minor modifications done due to the VLE beam result in slightly bigger spots at the experiment which is basically transparent to the experiment. Swapping between the "normal" secondary or tertiary beams to the VLE setup requires minor modifications that can be done during a short MD (8 h) time.

Following some initial discussions a request was made by the ATLAS collaboration to have the beam operational during the SPS-2003 run to mainly test the Tile Calorimeter in stand-alone mode, and later in 2004 to use it to test the LArgon em calorimeter as well as several tracking detectors during the combined run of a complete ATLAS wedge. Details of the VLE beam design and implementation are described in the following sections. Operational experience as well as first results from the 2003 and 2004 runs and comparison with beam simulation are also discussed.

## 2 VLE beam design and instrumentation

Following the schematic layout of the VLE Beam as shown in Figure 1 the design criteria of its main components are presented here.

Existing large aperture magnets were used for the beam: QPL and QPS quadrupoles having an aperture of 200 mm, and MBPL dipole magnets that offer a high field of $1.8\,\mathrm{T}$ with a large aperture of $140\,\mathrm{mm}$. The parameters of magnets used can be found in [3].

In order to achieve the required rates for the VLE beam, in particular for the production of low energy pions at $1\,\mathrm{GeV/c}$, the incoming secondary beam has to approach the maximum allowed intensity, $10^8$ particles per spill. The remaining secondary hadron beam downstream the T48 target is stopped at the dump build around the second dipole. The dump itself is thick enough to stop all the hadrons: 3.2 m along the beam line, 80 cm transverse of iron plus two 80 cm concrete blocks at the entrance and exit sides. However, muons come along with the secondary beam, typically $1\%$ of the pion flux within an area of $10 \times 10\,\mathrm{cm}^2$. The dump will help by multiple scattering these muons, therefore reducing their flux per $\mathrm{cm}^2$ at the detector.

The spectrometer around the fourth bending magnet is to measure the beam momentum particle-by-particle. It is required if the experiment wants to go beyond the beam momentum resolution defined by the collimator. The beam geometry, optics, and simulation is described in the standard files used by AB/ATB/EA that can be found in [3]. In the following sections a description of the major components of the beam is given.

### 2.1 The T48 target

The VLE beam is a tertiary beam, produced from a secondary target hit by a secondary medium energy beam in the range of $40$ to $100\,\mathrm{GeV/c}$. The target material and dimensions as well as the secondary beam momentum are chosen to optimize the flux of low energy particles. Its longitudinal position is determined by two factors: a) to have the shortest distance to the experiment in order to minimize the pion decay and b) the required space for the beam elements, in particular the large dipoles with their limited bending power.

Figure 2 shows the particle yield at the T4 target using a GEANT simulation for an incident proton beam at $400\,\mathrm{GeV/c}$. For the range of the secondary beams used here, a maximum yield of a few $10^{-5}$ hadrons (mixed of $\pi^+$ and protons) and $10^{-6}$ positrons per incident proton is expected. For a negatively charged secondary beam the same flux is expected for electrons or pions ($\pi^-$ in this case) but a negligible number of anti-protons. In general, positively charged hadron beams in the range of $40 - 100\,\mathrm{GeV/c}$ contain about 25-30% protons while negatively charged hadron beams are pure $\pi^-$s.

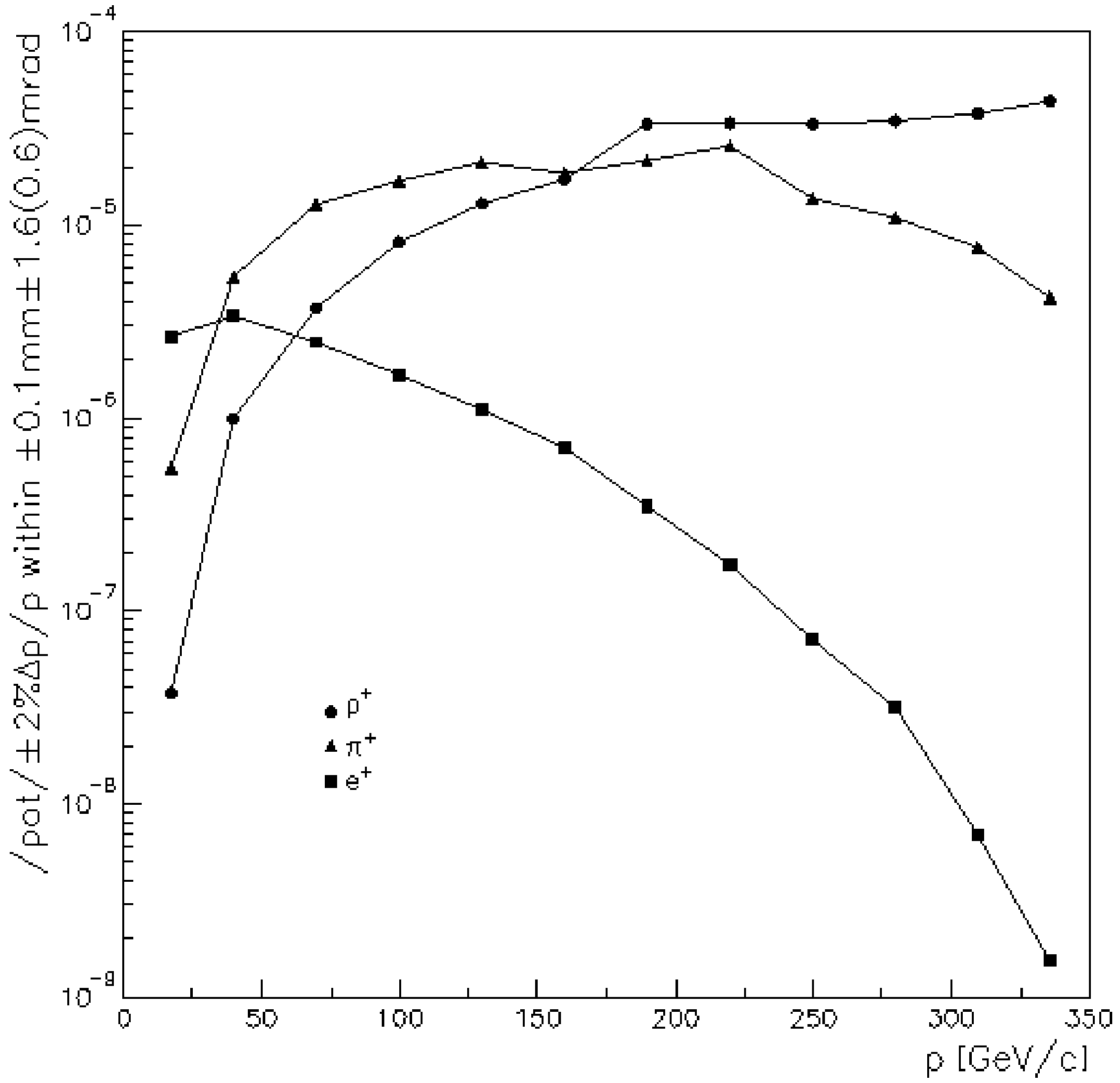

Figure 2: *Secondary particle production at the T4 target (*$300\,\mathrm{mm}$ *of Be) simulated with GEANT3 for an incident proton beam of* $400\,\mathrm{GeV/c}$*. The quoted rates are per incident proton within the full acceptance of the H8 beam line.*

The VLE particle production at the T48 target from the GEANT simulation is shown in Figure 3 for a $15\,\mathrm{cm}$ long lead target and secondary hadron (pion) beams of different momenta. According to the simulation, for an initial proton flux of $2 * 10^{12}$ protons per spill at the T4 target, about $10^3$ particles at $1\,\mathrm{GeV/c}$ are expected at the experiment, increased by an order of magnitude at $10\,\mathrm{GeV/c}$. However, as mentioned above, these low energy pions would have to be separated from a very high muon background that comes along with the high intensity secondary hadron beam, estimated to about $10^6$ per spill. Using a tertiary beam instead of a secondary (i.e start with a secondary beam out from T4 of $180\,\mathrm{GeV/c}$ and use the filter target to produce a tertiary beam of $40 - 60\,\mathrm{GeV/c}$ which is transported up to the T48 target,) would lower the muon background but at the same time would decrease the VLE beam flux by a factor 10 to 100 depending on the exact configuration. Therefore this option was not considered.

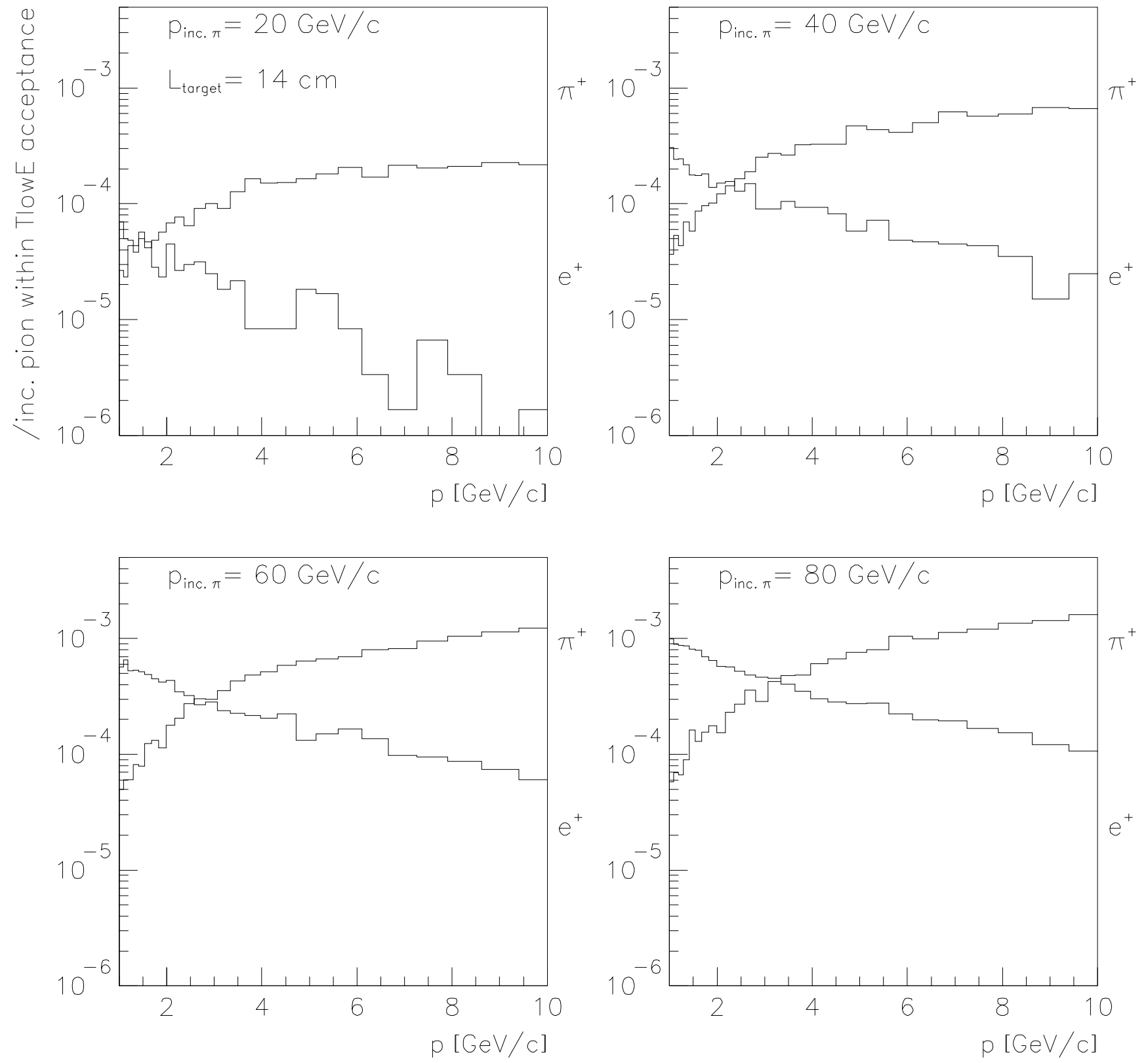

Figure 3: *Particle yield for the T48 target using* 15 cm *of lead, for several incident secondary pion beams - GEANT3 simulation.*

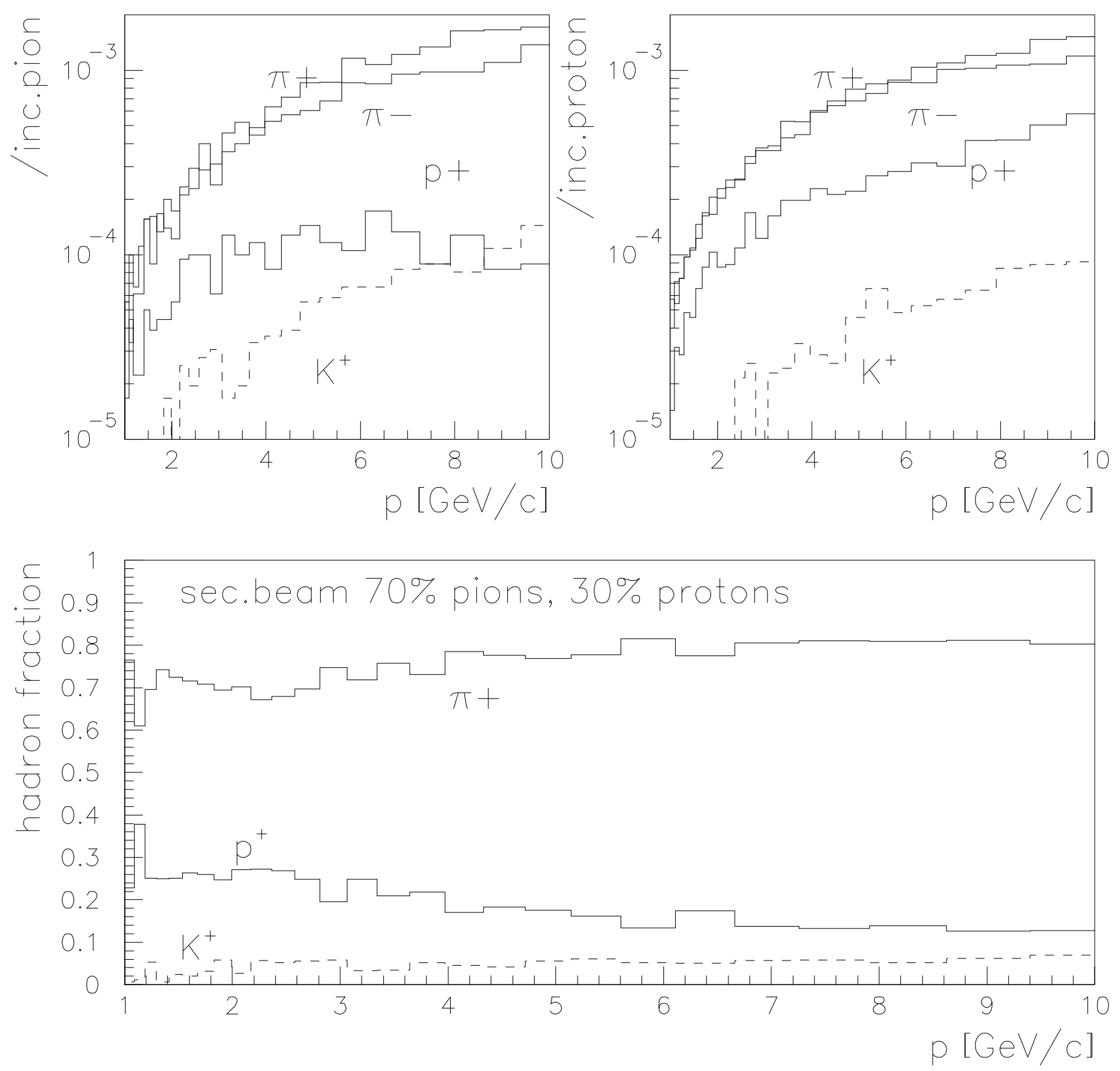

Figure 4: *Simulated (GEANT) hadron composition of the VLE hadron beam. The upper two plots show the production of tertiaries from a 15cm lead target per incident pion or proton. With the typical composition of the secondary beam of pion:proton 70:30, the VLE hadron beam is composed as indicated in the lower plot.*

The T48 target heads are put in a "standard" XCON frame, which can hold three different target heads, and can be completely retracted (out of the beam) to allow the passage of the high-energy secondary or tertiary beams when the VLE beam is not operational. The target heads can be selected remotely to be one of the following:

- polyethylene, $100\,\mathrm{cm}$
- lead, $0.5\,\mathrm{cm}$
- lead, $15\,\mathrm{cm}$

All three target heads are cylindrically shaped with a diameter of $4\,\mathrm{cm}$. The long lead target is optimized for pion production, the short lead target serves as an electron target. The polyethylene target produces an intentionally mixed beam, also at momenta close to 9 GeV/c. The detailed simulation results can be found in [3].

Having identified the electrons, the VLE hadron beam is composed of pions, protons and kaons. Figure 4 shows the simulated (GEANT) ratio of hadrons as a function of the VLE momentum. The production yield of VLE tertiaries from a secondary pion or proton is indicated in the upper plots. With a secondary beam momentum of 80 GeV/c the pion-proton ratio is about 70:30 (see figure 2), resulting in a VLE pion fraction of about 80 % in the whole VLE momentum range.

### 2.2 Beam acceptance and momentum selection

To maximize the beam acceptance, large aperture quadrupoles are used, with the first one located as close as possible to the T48 target, operated at its maximum strength. The beam optics are shown in Figure 5. The VLE beam acceptance is $3 \times 7\,\mathrm{mm\,mrad}$ in the horizontal and $3 \times 30\,\mathrm{mm\,mrad}$ in the vertical plane.

For the momentum selection the MBPL magnets are chosen since they offer a high aperture ($140\,\mathrm{mm}$) and a high bending power of about $4.0\,\mathrm{Tm}$ at maximum current. The deflection angle at the top momentum of $10\,\mathrm{GeV/c}$ is $120\,\mathrm{mrad}$ which determines the relative position between the first and the second bend and consequently of the overall length of the VLE beam line. However, due to existing constraints in the experimental hall, the distance between the bends had to be modified and finally the VLE beam was designed for a top energy of $9\,\mathrm{GeV/c}$ with a bending angle of $120\,\mathrm{mrad}$ at each MBPL. The total beam length from the T48 target to the ATLAS experimental setup in the H2A area is about $45\,\mathrm{m}$, which corresponds to about 85% of the decay length for $1\,\mathrm{GeV/c}$ pions.

It is important to have a good overlap between the energy range of the VLE beam and the existing high energy secondary and tertiary beams so that ATLAS can combine data from both beams, for example in linearity studies. Since a secondary electron beam down to $5\,\mathrm{GeV/c}$ with a reasonable rate can be produced at the primary target, they were passed through the VLE setup (of course without inserting the T48 target) in order to cross-calibrate the high energy and the very low energy momentum selection.

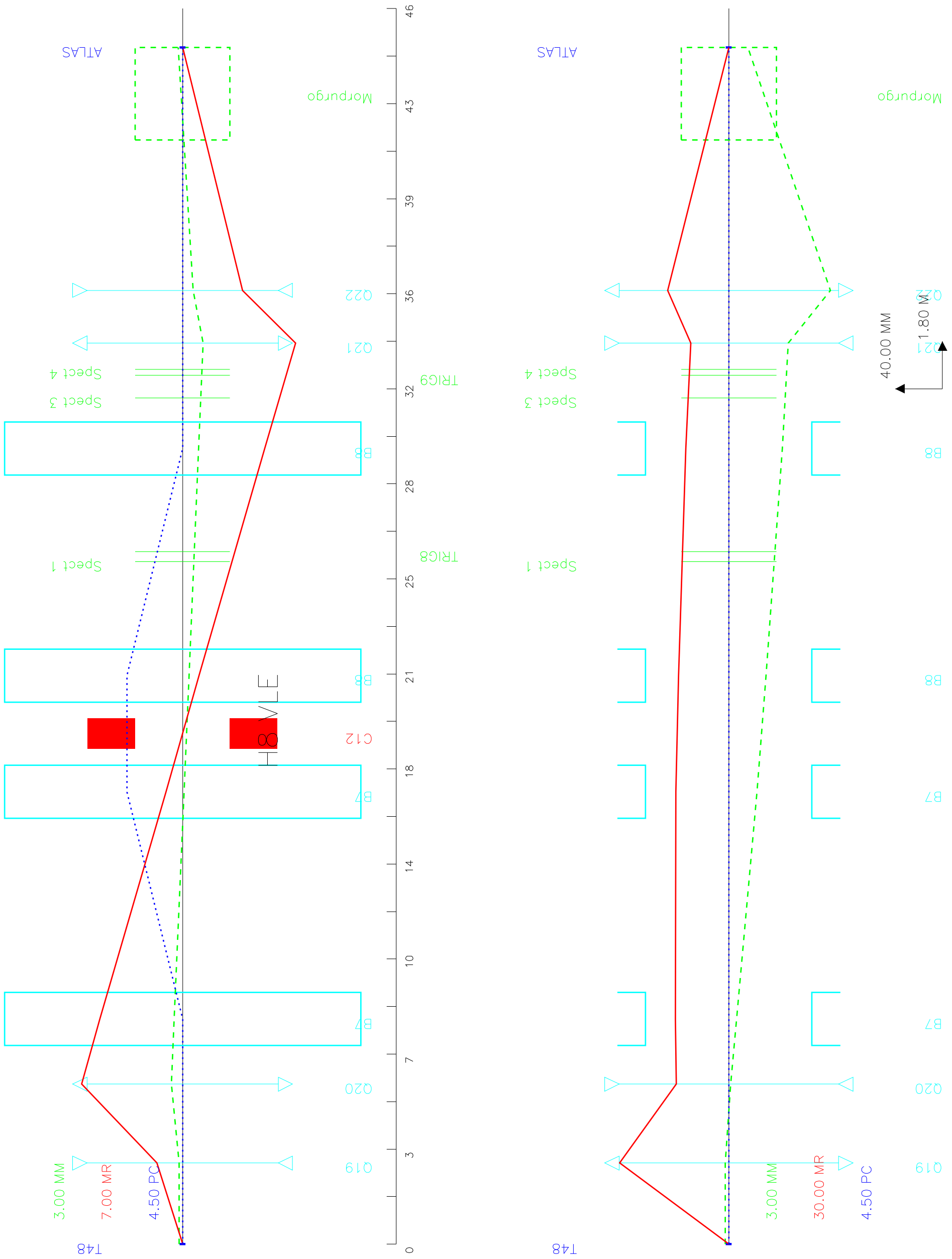

Figure 5: *The VLE beam optics diagram: horizontal (top) and vertical plane (bottom).*

The position of the third and fourth bends is a mirror image of the first two such that the beam is naturally brought back on the same line. The four MBPL magnets are powered in pairs, with the magnets within a pair connected in series: first and second MBPL as *BEND-7*, third and fourth MBPL as *BEND-8*. All the quadrupoles are individually powered.

### 2.3 Beam spectrometer

With the momentum defining collimator at full opening, the momentum acceptance of the VLE beam is 4.5%. Although such a resolution is quite acceptable in most cases, the beam line is equipped with a spectrometer that can be used to measure the momentum of individual particles. The spectrometer consists of three delay wire chambers located around the fourth bend as shown in Figure 6. Using the two chambers downstream of the magnet (Spect-3 and Spect-4 in the figure) and assuming the rotation point at the magnet center the angular deflection of each particle can be determined as

$$\theta = \cos^{-1} \frac{L_1(L_4 - L_3) + (\frac{x_3 L_4 - x_4 L_3}{L_4 - L3} - x_1 - L_1 \tan\theta_0)(x_4 - x_3)}{\sqrt{L_1^2 + (\frac{x_3 L_4 - x_4 L_3}{L_4 - L3} - x_1 - L_1 \tan\theta_0)^2}\sqrt{(L_4 - L_3)^2 + (x_4 - x_3)^2}} \quad (1)$$

where $L_1 = -4.235\,\mathrm{m}$, $L_3 = 1.942\,\mathrm{m}$, $L_4 = 3.022\,\mathrm{m}$ are the distances of the three spectrometer chambers as shown in the figure, $\theta_0$ is the nominal bending angle of $120\,\mathrm{mrad}$ and $x_1, x_3, x_4$ are the reconstructed coordinates. The chambers have an aperture of $100 \times 100\,\mathrm{mm}^2$ and a resolution of about $300\,\mu\mathrm{m}$, which gives an overall precision of $0.5\%$ for the individual particle measurement.

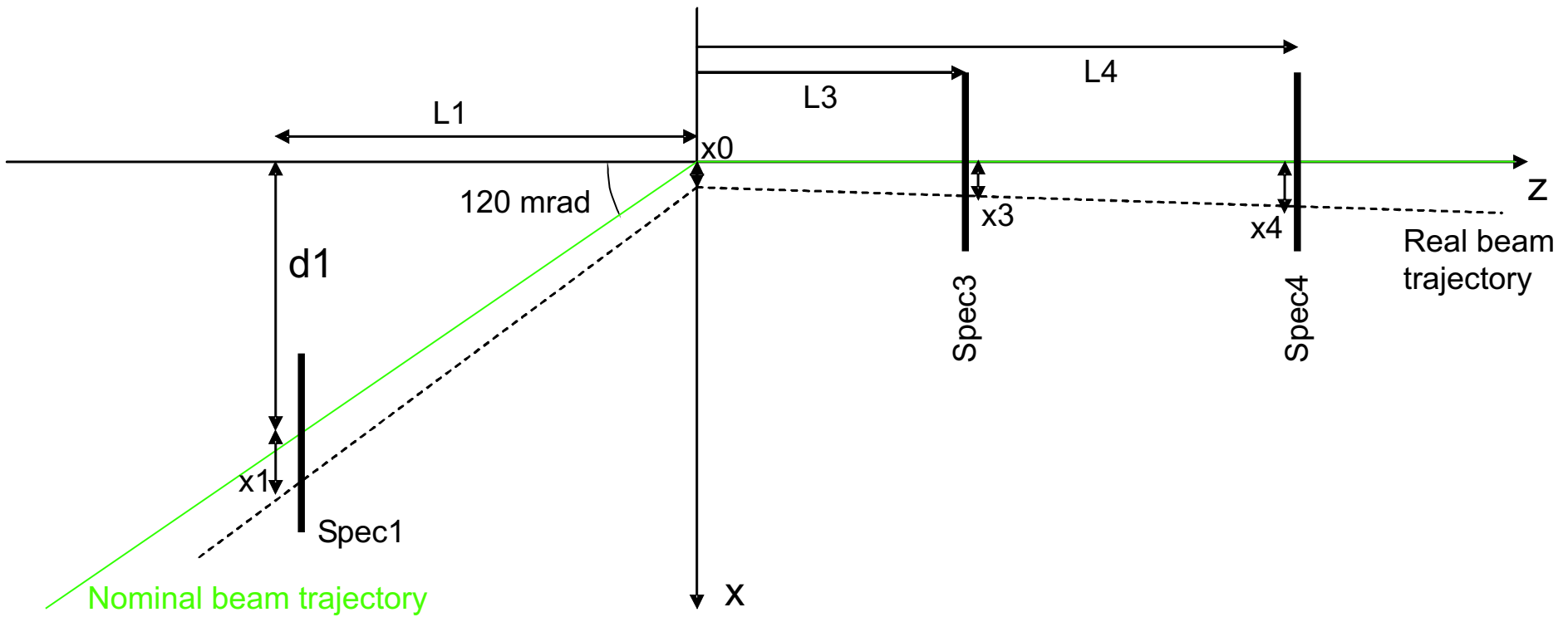

Figure 6: *Geometry of the VLE spectrometer for momentum analysis.*

From the angular deflection the momentum of the particle can be determined from the well known relation:

$$p\,[\mathrm{GeV/c}] = \frac{299.79}{\theta\,[\mathrm{mrad}]} \times \int Bd\ell\,[\mathrm{Tm}] \tag{2}$$

where $\int Bd\ell$ is the field integral of the magnet that corresponds to the current value set. For the case of electrons, the energy loss due to synchrotron radiation in the dipole field has to be taken into account. For a $10\,\mathrm{GeV/c}$ electron it corresponds to about $1\ \mathrm{MeV}$ or $0.01\%$.

### 2.4 Particle identification

For the particle identification, i.e. electron to pion (hadron) separation, a threshold Cherenkov counter is installed in the space between the third and fourth bend upstream of the first spectrometer chamber (see figure 1). The counter has an active length of $2.5\ \mathrm{m}$ and is filled with He gas. In Figure 7 the threshold pressure for different particles as a function of the momentum is shown, indicating that $e/\pi$ separation is rather easy at the full VLE beam momentum range. Using the PDG formula for a $2\ \mathrm{m}$ counter, set at pion threshold we should expect about 340 photoelectrons for electrons at $1\ \mathrm{GeV/c}$ which should be easy to detect, even after dividing by a factor of two due to different efficiencies, etc.

## 3 VLE beam performance

The commissioning of the VLE beam was performed during the third week in August 2003. The beam was delivered to a common setup of the ATLAS Transition Radiation Detector (TRT) and barrel hadronic Tile Calorimeter (TileCal).

In Figure 8 the beam profiles measured at the spectrometer chambers are shown. At the experiment, after the two QPS magnets providing the final focusing, the beam is made almost parallel to match to the size of the experimental trigger counters of $3 \times 3\,\mathrm{cm}^2$.

Using the data from the spectrometer the beam momentum of individual particles was calculated, as shown in in figure 9. The width of the distributions increases with increasing mean momenta being defined by the collimator slit to be $\Delta p/p = \pm 4.5\%$. All nine spectra are recorded with the collimator fully open. For the $9\,\mathrm{GeV/c}$ case a scan of the momentum defining collimator was made, with its jaws closed to half width, once located on the positive and once on the negative side. The result is shown in the lower right plot where, as expected, the mean momentum is shifted to a lower and respectively higher value.

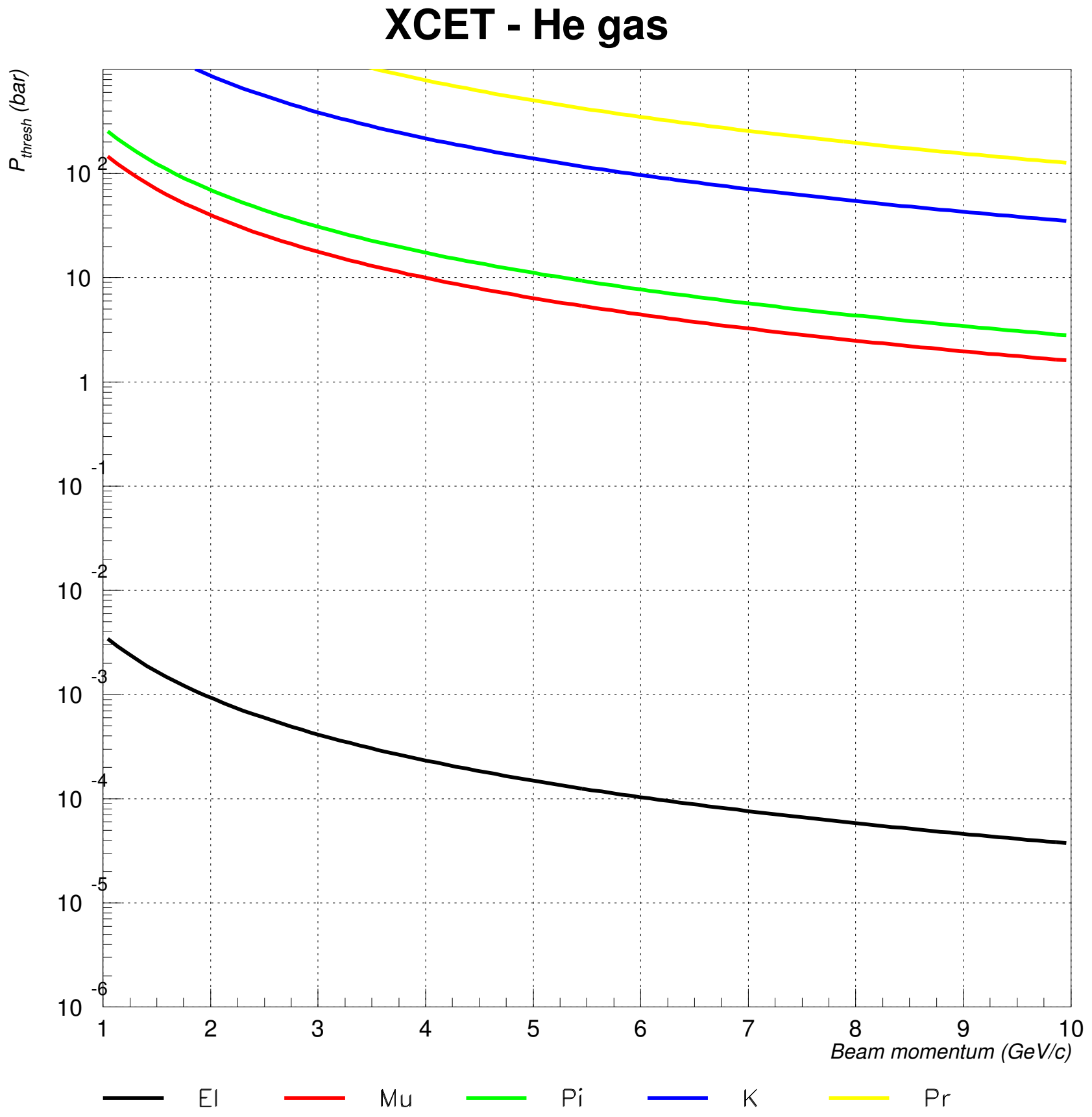

Figure 7: *Pressure threshold versus beam momentum for different particles for a He gas counter.*

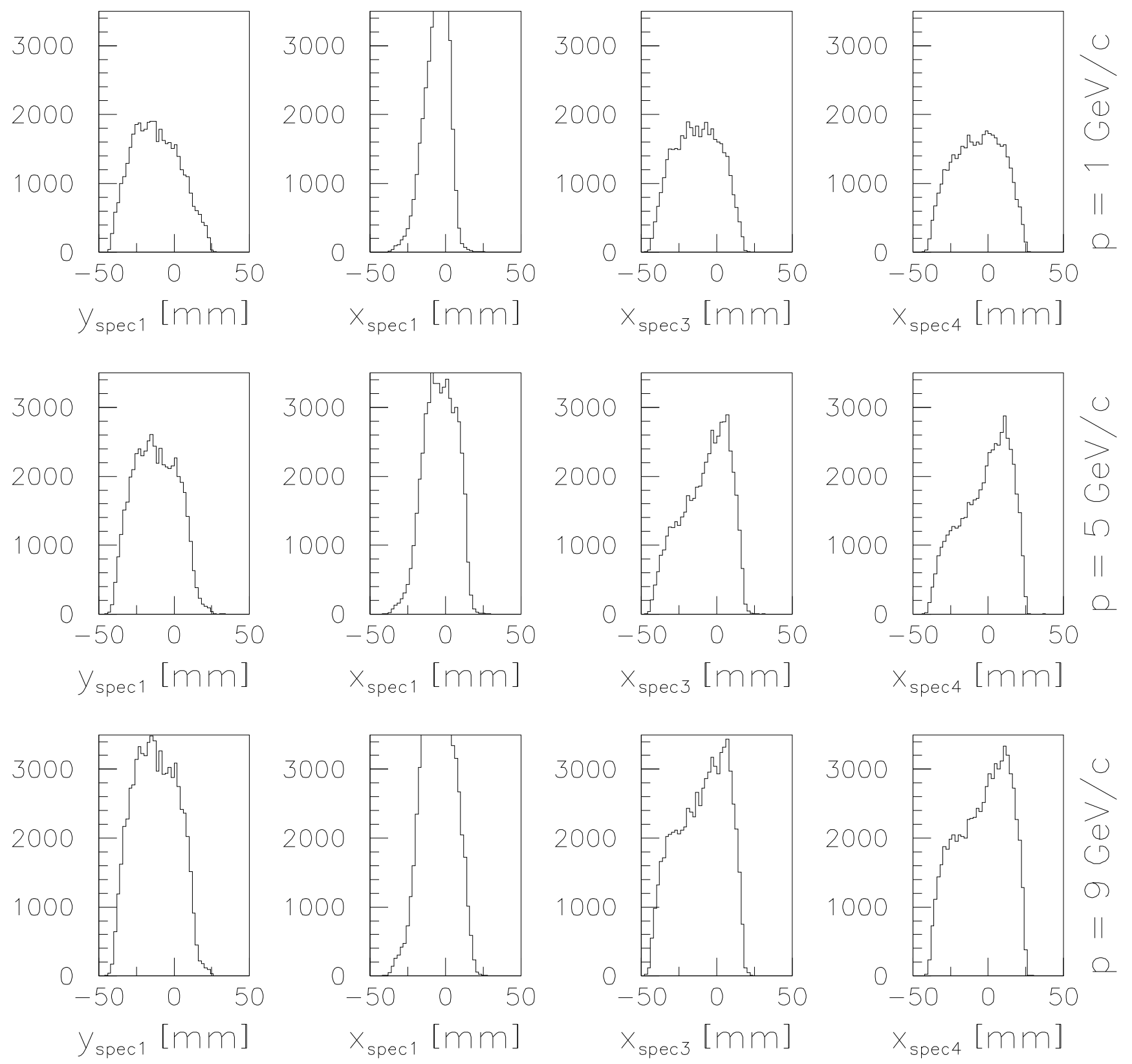

Figure 8: *VLE beam profiles measured by the spectrometer chambers for three beam momenta.*

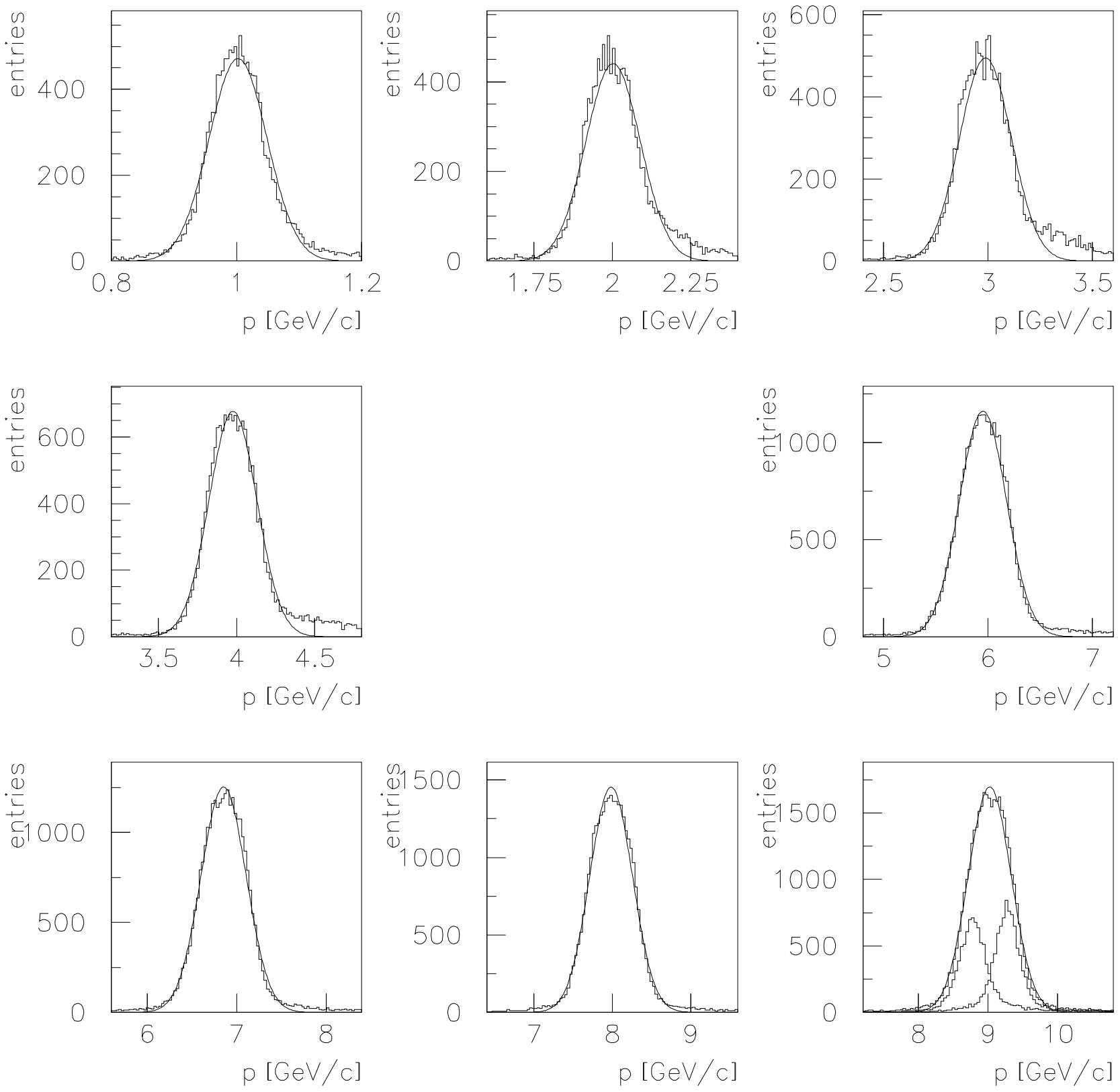

Figure 9: *Single particle momentum reconstruction for the VLE beam. For the* $9\,\mathrm{GeV/c}$ *case the separate profiles corresponding to different settings of the momentum selection collimator are shown: left:* $[-30, 0]\,\mathrm{mm}$, *right:* $[0, +30]\,\mathrm{mm}$, *center:* $\pm 40\,\mathrm{mm}$

The peak and width of the beam momentum spectra using a Gaussian fit versus the set momentum is shown in figure 10.

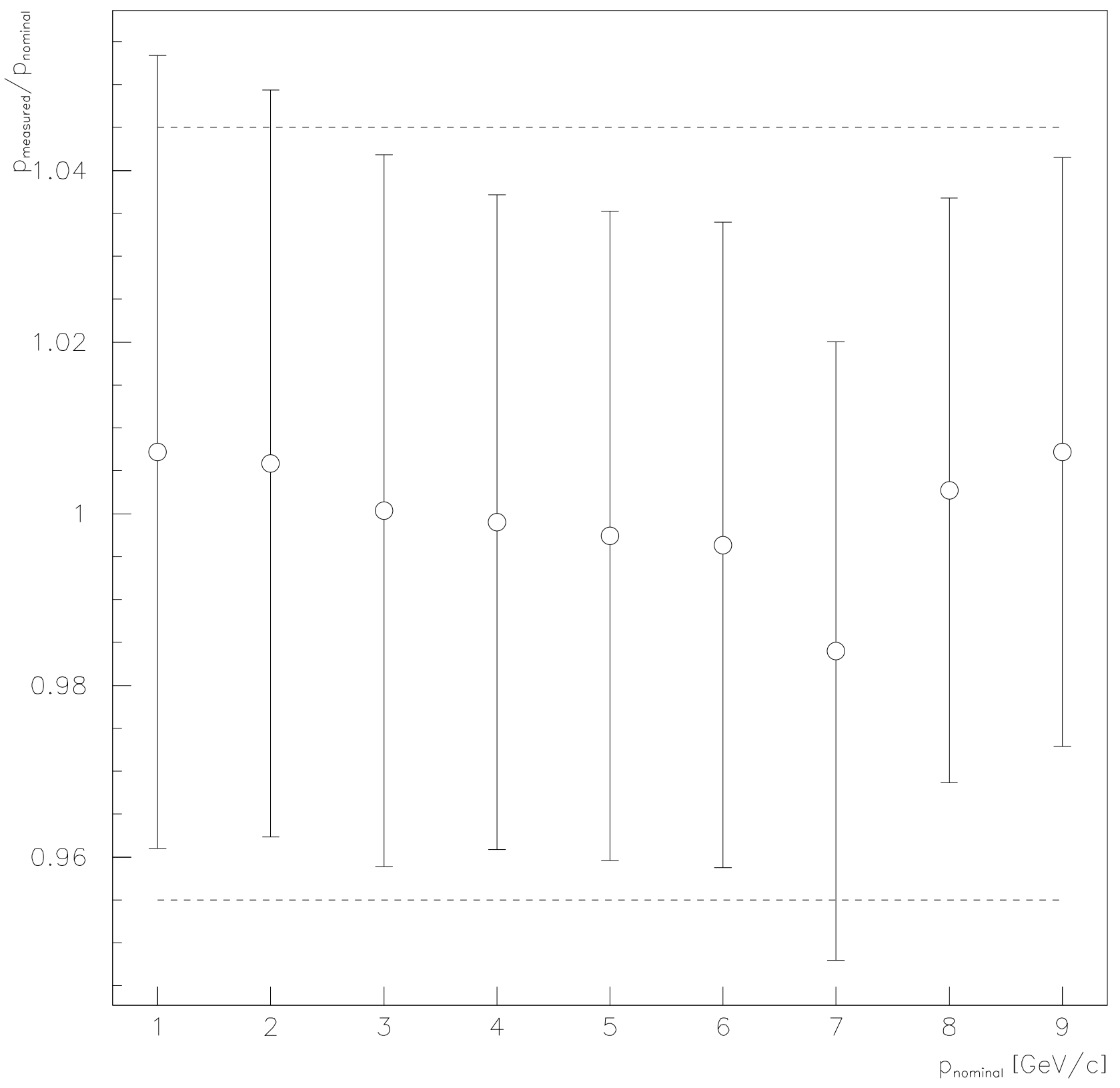

Figure 10: *Relative VLE beam momentum. The error bars correspond to the beam momentum width and the horizontal lines indicate the* $\pm 4.5\,\%$ *acceptance of the beam.*

Depending on the type of the secondary beam and the T48 target head chosen, the VLE beam can be set for electron or pion mode. Electron beams are easier to produce, are pure and intensities of at least 1 kHz can be easily obtained. On the contrary, hadron (pion) beams have much lower rate and always a non negligible electron and muon background. In figure 11 the total VLE particle rate measured by the coincidence of two $10 \times 10\,\mathrm{cm}^2$ scintillator counters located next to spectrometer chambers 1 and 4 (see figure 1) is shown for an incident hadron beam of $80\,\mathrm{GeV/c}$ at maximum allowed intensity ($\sim$ few $\times\, 10^6$ ppp). A good agreement

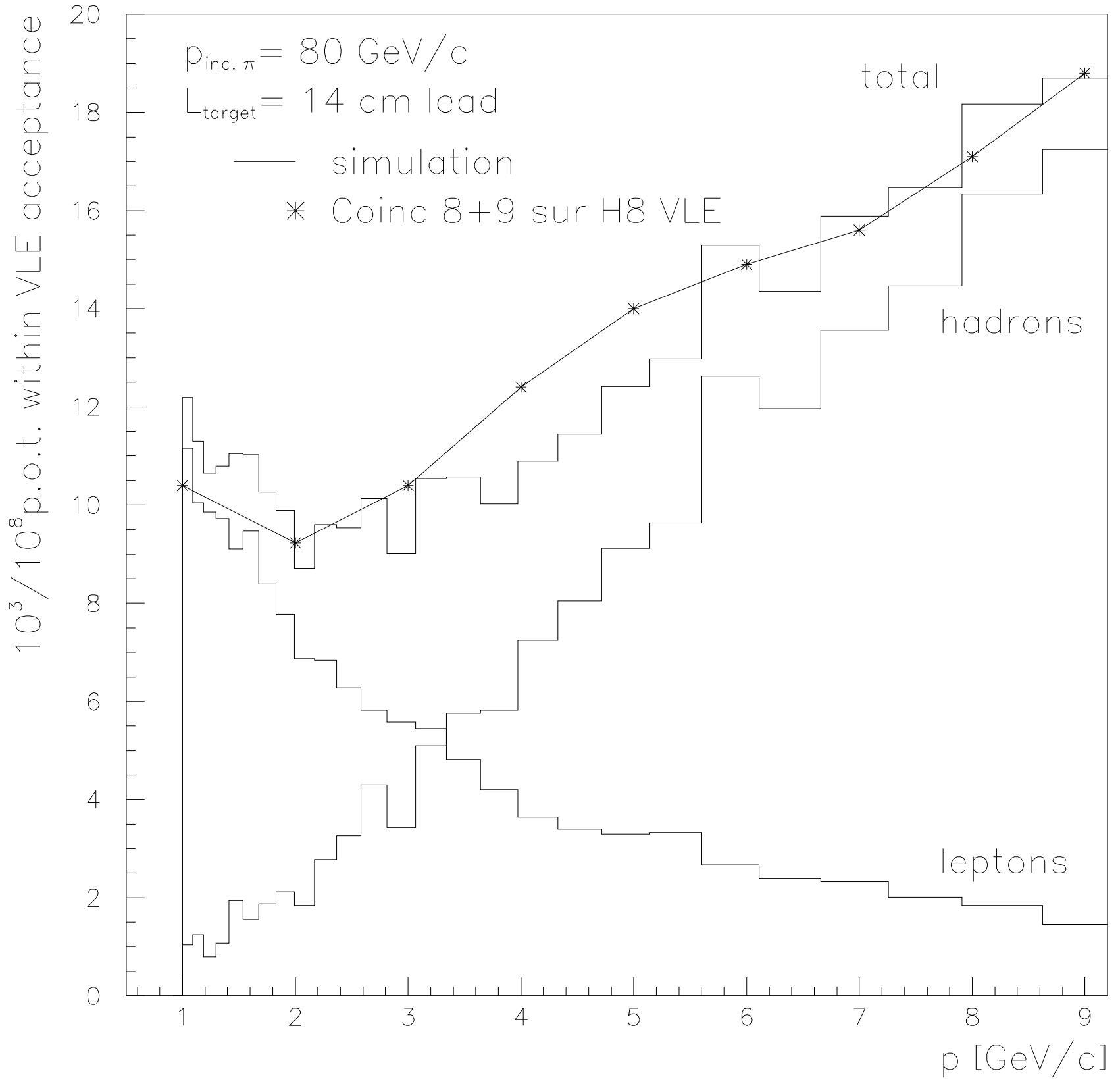

Figure 11: *VLE beam particle rate per spill (*$4.8\,\mathrm{s}$ *for an incident hadron beam of* $80\,\mathrm{GeV/c}$ *with the T48 target set to* $15\,\mathrm{cm}$ *of lead).*

with the GEANT simulation is observed.

Due to time constraints the threshold Cherenkov counter was not used during the test. However, using the ATLAS Tile-Calorimeter, a preliminary estimate of the electron content in the VLE pion beam was made, which are shown in figure 12. Particle rates/spill observed by the ATLAS detector were always an order of magnitude lower than using the triggers of the VLE spectrometer. This is due to small size of triggers used by ATLAS TileCal covering only a part of the total beam size. Pion production is high at $9\,\mathrm{GeV/c}$ and decreases rapidly with lower nominal momentum. The number of particles/spill in pion mode is indicated in Figure 11, where Figure 12 shows the pion content as a function of the momentum. At $1\,\mathrm{GeV/c}$ the relative rate of pions is down to $10\%$, which corresponds to a few hundred/spill.

Due to pion decay, a pion (hadron) beam is always accompanied by a muon halo. In order to have the highest possible rate for pions in the VLE beam, the incoming secondary hadron beam had to have the highest allowed rate, which implied a rate of $\sim\ 10^6$ muons per spill, even downstream of the VLE dump. However, their impact on the ATLAS detectors, TRT and Tile-Calorimeter, was minimal. A large part of the muon background was rejected using a muon veto counter of about $20 \times 20\,\mathrm{cm}^2$ located on the straight beam axis next to spectrometer 1 chamber (see figure 1). Moreover, the incoming muons from the secondary beam would successively see the field of the two bends on the beam axis that are of the same polarity, therefore making a non negligible total deflection in the horizontal plane, even at $9\,\mathrm{GeV/c}$. In figure 13, a simulation shows the displacement of the muon background profile at the front face of the Tile-Calorimeter in H8A and in H8B at the entrance to the MBPL. With the VLE beam set for $9\,\mathrm{GeV/c}$ the deflection of the muon background spot is about $0.7\,\mathrm{m}$, which using the fine segmentation of the calorimeter, is sufficient to separate it. However, for the lowest VLE momentum of $1\,\mathrm{GeV/c}$ the separation is only few cm, therefore only the efficiency of veto counters and punch-through detectors can help.

Another source of non-reducible muon background comes from the pion decays in the length of the VLE beam. These muons can only be tagged using the calorimeter segmentation, which of course is not ideal depending on the type of measurement performed.

## 4 Impact on high energy beam performance

The installation of the H8 VLE beam setup required minor modifications to the elements serving the high energy beam on the straight line. This concerns mainly a pair of quadrupoles resulting in a slight decrease of the focusing strength. In figure 14, the beam profile of the straight beam after the modifications due to the

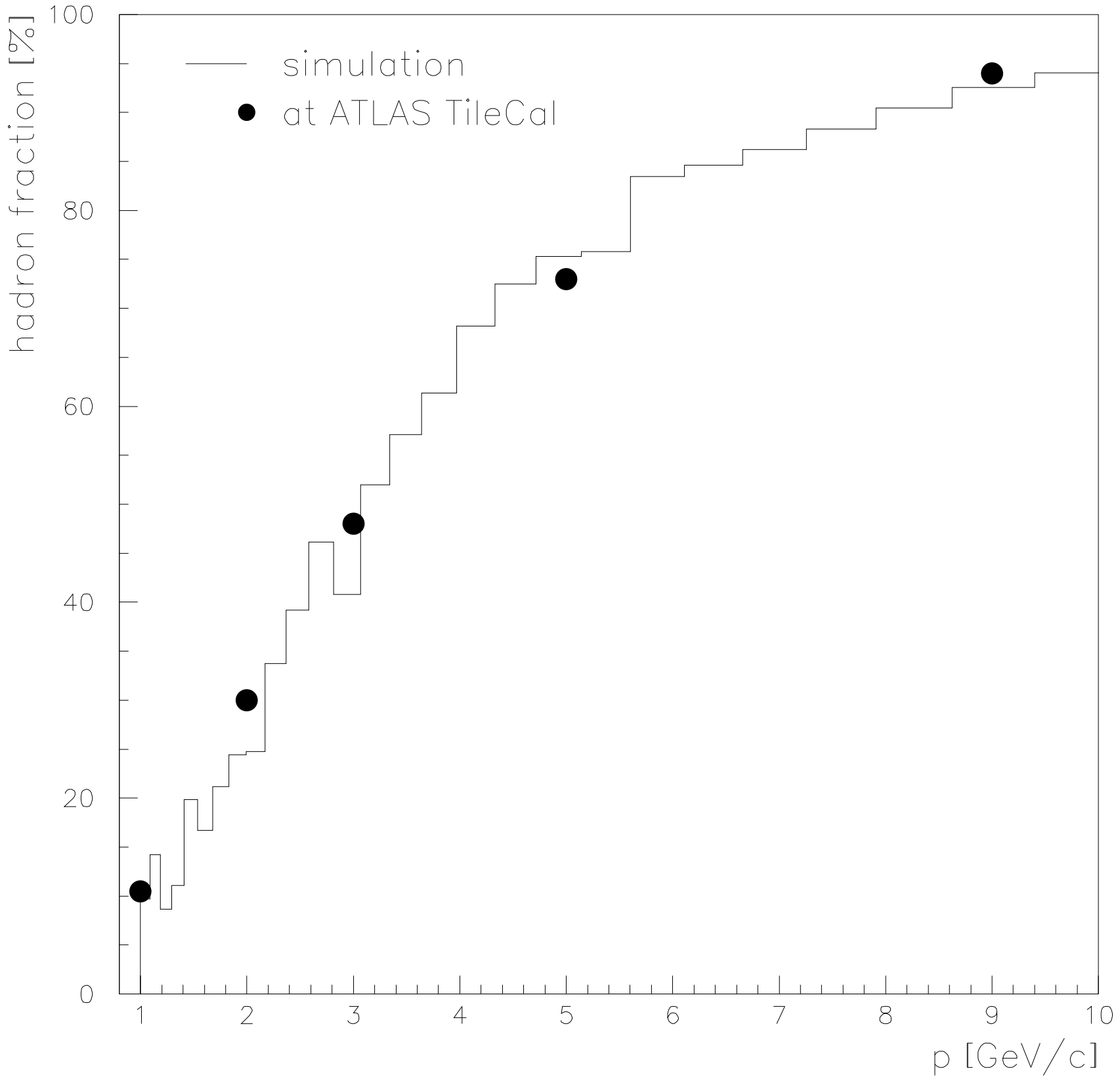

Figure 12: *Pion content of the VLE hadron(pion) beam as a function of the beam momentum, as measured by the ATLAS Tile Calorimeter (courtesy B.Stanek et al., Atlas TileCal). The remaining fraction of the beam is mainly electrons produced at the T48 target and transported up to the experiment.*

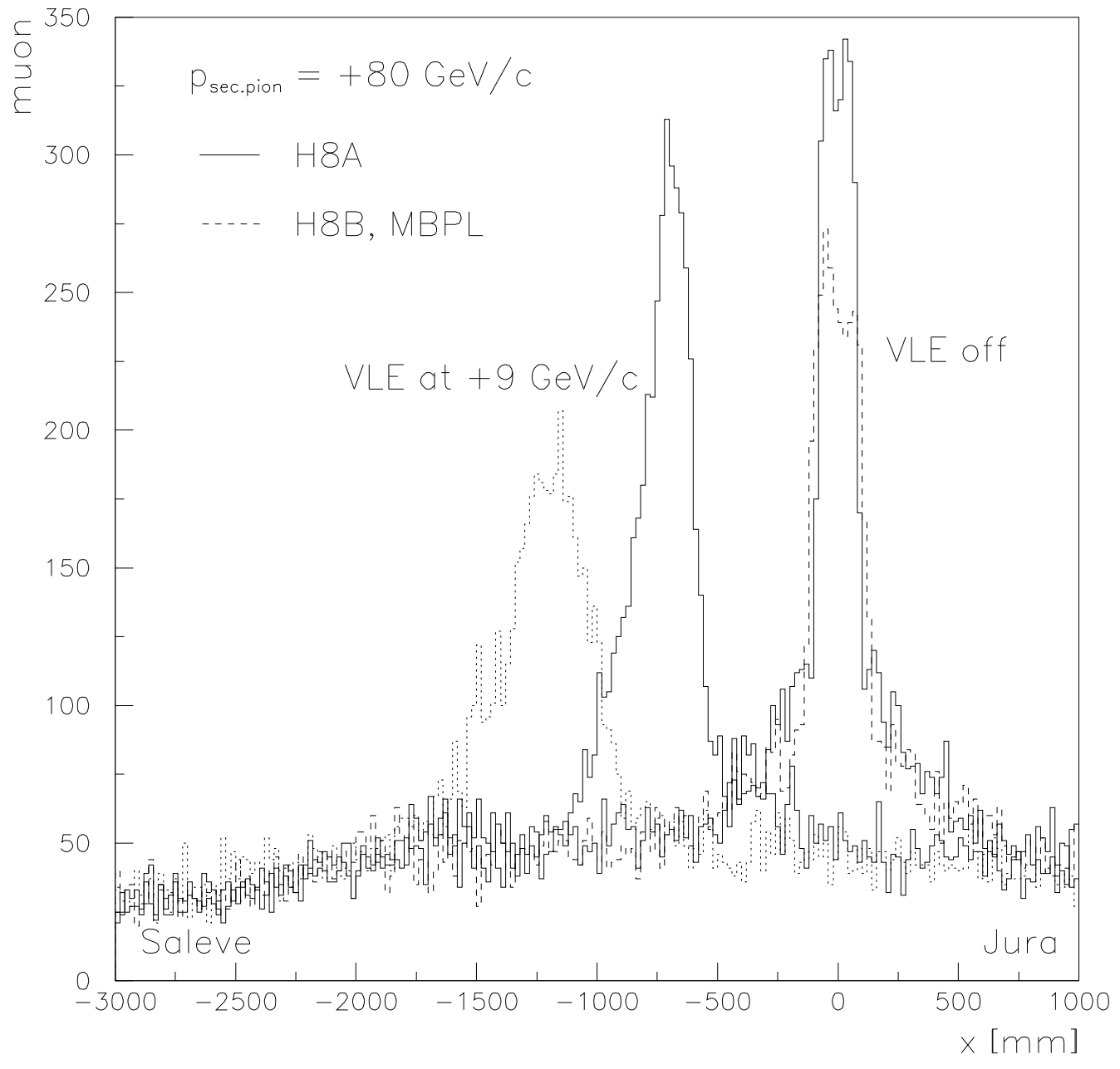

Figure 13: *Profile of the muon background from the incoming secondary beam onto the calorimeter front face and in the ATLAS muon area, with the VLE beam set for* $9\,\mathrm{GeV/c}$ *particles.*

VLE setup is shown. No significant differences are observed with respect to the previous situation, only some slight broadening of the spectrum as expected.

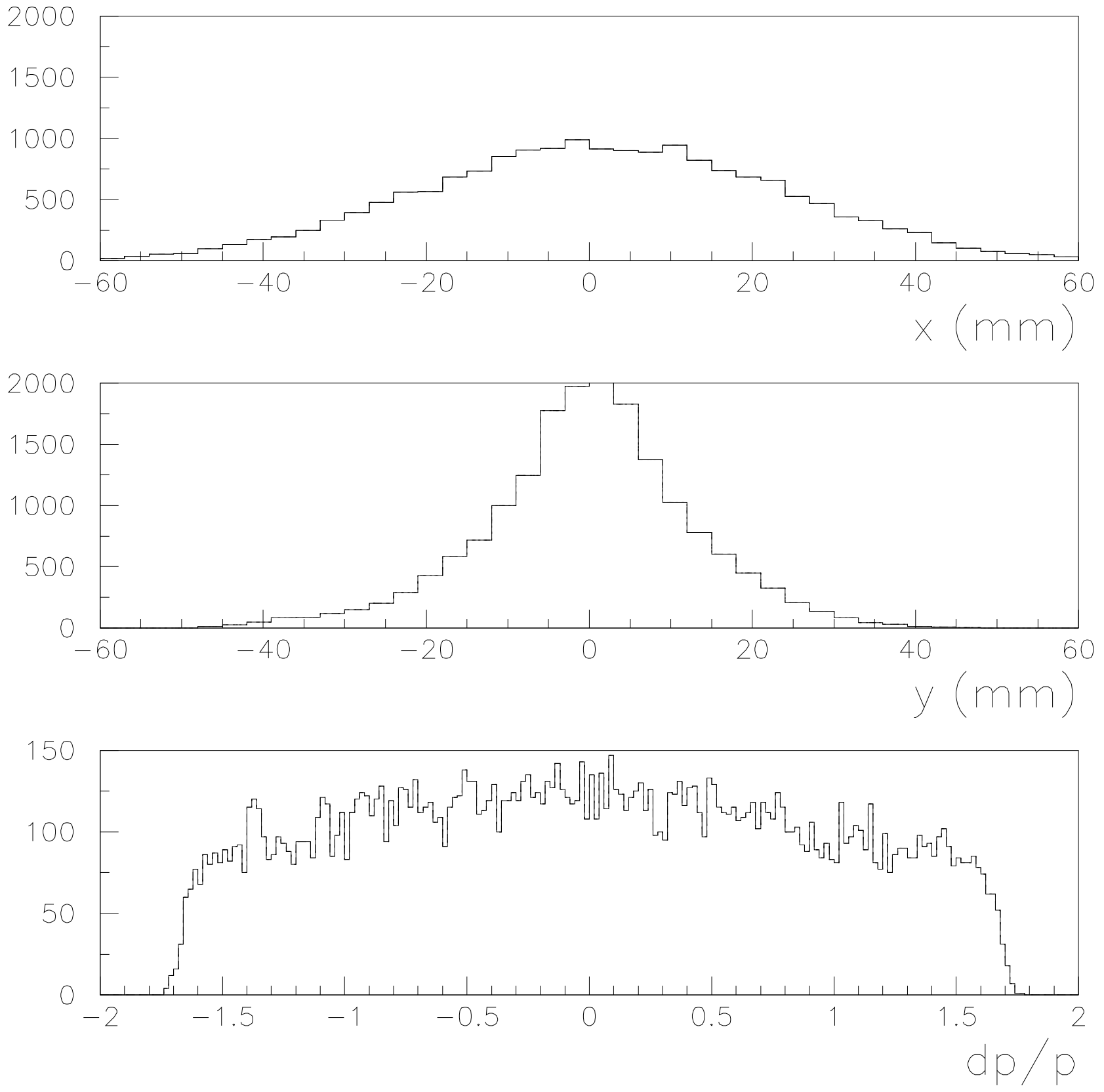

Figure 14: *Beam spot for* $p > 10\,\mathrm{GeV/c}$ *beam at H8A for full acceptance.*

# 5 Summary

The design and performance of the new Very Low Energy extension of the H8 beam line of SPS was presented. The new set-up can provide hadron (pion) and electron beams of both positive and negative polarities to the ATLAS detector setup in H8A area. From the first tests and operation in August 2003 and during 2004 the beam performance was within expectations, matching well the needs of the ATLAS collaboration.

With this extension the H8 beam line can provide beams from $1\,\mathrm{GeV/c}$ up to the top SPS energy of $400\,\mathrm{GeV/c}$, thus making it a unique facility for the performance tests of the ATLAS detectors.A similar VLE setup was installed for the CMS collaboration on the H2 beam line operational since 2004.

The authors would like to thank the colleagues of the AB/ATB-EA group who helped in the project. Particular thanks to N. Doble and K. Elsener for providing the basic design and first studies respectively.